\documentclass[prl,amsmath,amssymb,twocolumn]{revtex4-1}
\usepackage{graphicx}
\usepackage{subfigure}
\usepackage{adjustbox}
\usepackage{bm}
\usepackage{color}
\usepackage{braket}
\usepackage{standalone}
\usepackage{multirow}
\usepackage{tikz}
\usepackage{mathrsfs}
\usepackage{dsfont}

\usepackage[colorlinks,bookmarks=true,citecolor=blue,linkcolor=red,urlcolor=blue]{hyperref}

\newcommand{\bea}{\begin{eqnarray}}
\newcommand{\eea}{\end{eqnarray}}

\begin{document}

\title{Particle-Hole Duality, Emergent Fermi Liquids and Fractional Chern Insulators\\ in Moir\'e Flatbands}

\author{Ahmed Abouelkomsan$^1$, Zhao Liu$^2$}
\email{zhaol@zju.edu.cn}
\author{Emil J. Bergholtz$^1$}\email{emil.bergholtz@fysik.su.se}
\affiliation{$^1$Department of Physics, Stockholm University, AlbaNova University Center, 106 91 Stockholm, Sweden\\
$^2$Zhejiang Institute of Modern Physics, Zhejiang University, Hangzhou 310027, China}
\date{\today}

\begin{abstract} 
Moir\'e flatbands, occurring, e.g., in twisted bilayer graphene at magic angles, have attracted ample interest due to their high degree of experimental tunability and the intriguing possibility of generating novel strongly interacting phases. Here we consider the core problem of Coulomb interactions within fractionally filled spin and valley polarized Moir\'e flatbands and demonstrate that the dual description in terms of holes, which acquire a nontrivial hole dispersion, provides key physical intuition and enables the use of standard perturbative techniques for this strongly correlated problem. In experimentally relevant examples such as ABC stacked trilayer and twisted bilayer graphene aligned with boron nitride, it leads to emergent interaction-driven Fermi liquid states at electronic filling fractions down to around $1/3$ and $2/3$ respectively. At even lower filling fractions, the electron density still faithfully tracks the single-hole dispersion while exhibiting distinct non-Fermi liquid behavior. Most saliently, we provide microscopic evidence that high temperature fractional Chern insulators can form in twisted bilayer graphene aligned with hexagonal boron nitride.
\end{abstract}

\maketitle

{\it Introduction.-}
Moir\'e superlattice systems have attracted enormous interest following the discovery of unconventional superconductivity and correlated states in twisted bilayer graphene near the magic angle \cite{cao2018unconventional,cao2018correlated,yankowitz2019tuning,Lu2019,spectrobilayer,origintwistedbilayer}. Owing to experimental advances in manufacturing van der Waals heterostructures \cite{vanderwaals}, there have been an increasing number of experiments investigating a plethora of novel strongly correlated two-dimensional systems with Moir\'e patterns. A salient example is bilayer graphene at a magic twist angle \cite{twistedbilayermodel1,twistedbilayermodel2,twistedbilayermodel3,alexpaper,futwisted} aligned with hexagonal boron nitride which gaps out the protected Dirac points of the graphene and the flatbands around charge neutrality acquire nonzero Chern numbers \cite{originofhbngap1,originofhbngap2,originofhbngap3,originofhbngap4}. Here, recent experiments show both emergent ferromagnetism \cite{Sharpe605} and a quantized anomalous Hall effect~\cite{serlin2019intrinsic} at different integer fillings suggesting that the underlying state is a Chern insulator. 

A related platform for studying correlated physics is the Moir\'e superlattice formed when trilayer graphene is aligned with boron nitride. Unlike magic angle twisted bilayer graphene, the bandwidth and the topology of the minibands can be controlled by the strength of the applied displacement field resulting in gate-tunable Mott insulating and superconducting states \cite{chen2019evidence,Chen2019}. Also in this case, a correlated ferromagnetic Chern insulator has been reported at a filling corresponding to one hole per Moir\'e unit cell \cite{chen2019tunable}. 

While a tremendous amount of attention has been paid to the aforementioned Moir\'e flatband systems, the detailed microscopic understanding of the crucial impact of Coulomb interactions has essentially been limited to investigating the spin and valley polarization at integer fillings of Moir\'e flatbands~\cite{repellin2019ferromagnetism}.

In sharp contrast, in this Letter, we provide a detailed microscopic study of Coulomb interactions within a fractionally filled spin-polarized Moir\'e flatband near one valley relevant for the experimental situation at finite doping. Motivated by the aforementioned recent experiments we focus on two particularly intriguing models. The first being ABC stacked trilayer graphene aligned with boron nitride which we refer to as TLG-hBN \cite{PhysRevB.99.075127,PhysRevLett.122.016401,senthilbridging,Constantinprb,PhysRevB.82.035409}. Here we take our fractionally filled band to be the valence band with Chern number $C =3$ and the valence band with Chern number $C = 0 $ obtained by switching the sign of the applied voltage. The second model describes twisted bilayer graphene aligned with boron nitride \cite{2019arXiv190108209Z} which we refer to as TBG-hBN. The active band is taken to be the valence band with Chern number $C =1$.

Based on the glaring analogy with Landau levels, it has been suggested that exotic correlated states such as fractional Chern insulators (FCIs) may quite generically form in $C\neq 0$ Moir\'e flatbands \cite{PhysRevB.99.075127}. Remarkably, however, we find that a large portion of the phase diagram at fractional filling in each of these bands is dominated by particle-hole symmetry breaking terms that are absent in Landau levels: rewriting the interaction Hamiltonian in terms of holes yields a dual formulation whereby the strongly interacting electron problem is converted into an effectively weakly interacting hole problem that is amenable to standard perturbative approaches. This results in a new paradigm for Fermi liquid states, with telltale experimental signatures, emerging from projected interactions alone. 
Moreover, our microscopic calculations also indicate instabilities at low electronic filling where the interactions, notwithstanding the paramount impact of the hole dispersion, lead to genuinely strongly correlated states including a high-temperature fractional Chern insulator.

{\em Setup.-}
Motivated by the recent discoveries \cite{Sharpe605,serlin2019intrinsic,chen2019evidence,chen2019tunable} along with theoretical predictions \cite{repellin2019ferromagnetism,zalatelanamolous,ladoprl} of spin and valley polarized insulators at integer fillings in various Moir\'e bands, we consider electrons with polarized spin and valley degrees of freedom interacting via the screened Coulomb potential
\begin{equation}
H_{a,\sigma}=\frac{1}{2} \sum_{\mathbf{q} \in R^2}  V(\mathbf{q})  :\rho_{a,\sigma}(\mathbf{q})\rho_{a,\sigma}(-\mathbf{q}):,
\label{eq1}
\end{equation}
where $\rho_{a\sigma}(\mathbf{q})$ is the density operator in valley $a$ with spin $\sigma$. We choose the Yukawa potential $V({\bf q})=\frac{e^2}{4\pi\epsilon S}\frac{2\pi}{\sqrt{|{\bf q}|^2+\kappa^2}}$ to describe the screening of the Coulomb interaction, where $e$ is the electron charge, $\epsilon$ is the dielectric constant of the material, $S$ is the area of the Moir\'e superlattice, and $\kappa$ measures the screening strength~\cite{supple}. While the full interaction Hamiltonian has $ U(2) \times U(2) $ symmetry including separate charge and spin conservation within each valley, we in Eq. (\ref{eq1}) neglect weaker terms like intervalley scattering and Hund's coupling which would break this symmetry.

When the electrons partially fill a Moir\'e flatband in valley $a$ with spin $\sigma$ while all other bands are either empty or completely filled, it is fair to project the interaction (\ref{eq1}) to the active flatband and neglect the band dispersion, so that the underlying physics driven by strong interactions is emphasized and disentangled from the quantitatively small effect of a remnant band dispersion. The projected Hamiltonian has the form
\begin{eqnarray}
\label{onesector}
H_{a,\sigma}^{\rm proj} = \sum_{\{\mathbf{k_i}\}} V^{a,\sigma}_{\mathbf{k}_1\mathbf{k}_2\mathbf{k}_3\mathbf{k}_4} c_{a,\sigma}^\dagger(\mathbf{k}_1) c_{a,\sigma}^\dagger(\mathbf{k}_2) c_{a,\sigma}(\mathbf{k}_3) c_{a,\sigma}(\mathbf{k}_4),\nonumber\\
\end{eqnarray} 
where $c^\dagger_{a,\sigma}(\mathbf{k})$ ($c_{a,\sigma}(\mathbf{k})$) creates (annihilates) an electron with momentum $\mathbf{k}$ and spin $\sigma$ in the partially filled flatband in valley $a$, and all ${\bf k}_i$'s are in the Moir\'e Brillouin zone (MBZ). The matrix element $V^{a,\sigma}_{\mathbf{k}_1\mathbf{k}_2\mathbf{k}_3\mathbf{k}_4}$ can be derived based on the effective model of the pertinent Moir\'e superlattice as detailed in the Supplemental Material \cite{supple}.  In what follows, we impose periodic boundary conditions on finite Moir\'e superlattices and use extensive exact diagonalization to study the low-energy properties of the Hamiltonian (\ref{onesector}) at various filling factors $\nu = N / (N_1 N_2)$, where $N$ is the number of spin-$\sigma$ electrons in the active band of valley $a$ and $N_1$ and $N_2$ are the number of unit cells in the two basic directions ${\bf a}_1$ and ${\bf a}_2$ of the Moir\'e superlattice. We choose $\frac{1}{2}\frac{e^2}{4\pi\epsilon a_M}$ as the energy unit, where $a_M$ is the lattice constant of the Moir\'e superlattice. Unless otherwise stated we set the Coulomb screening strength as $\kappa=1/a_M$.

\begin{figure}[t!]
	\includegraphics[width=1.08\columnwidth]{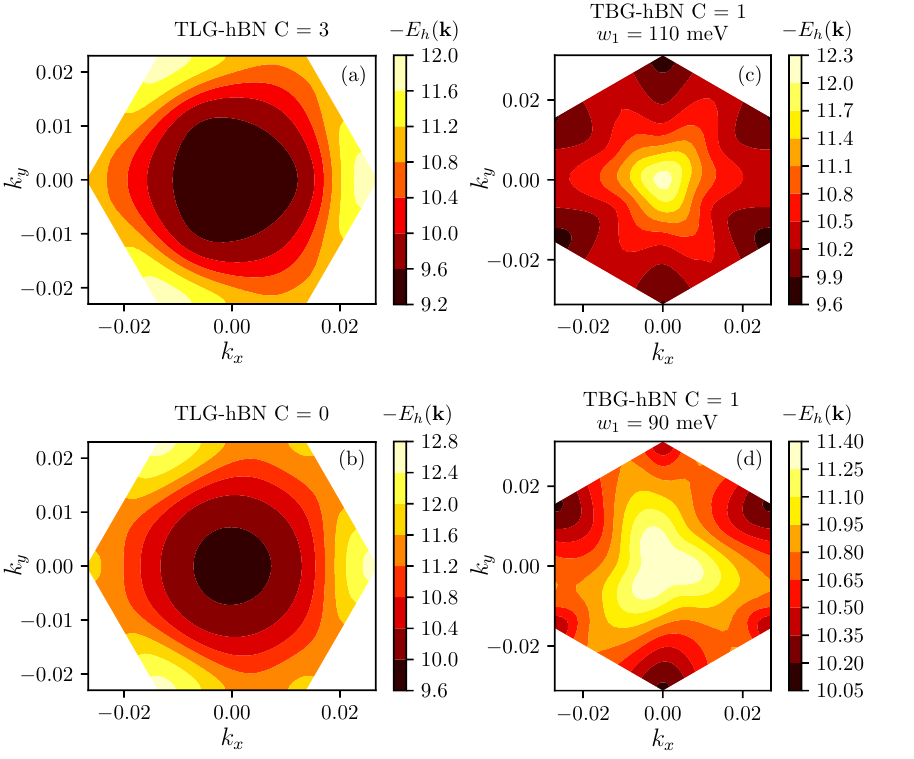}
	\caption{Contour plots of the single-hole dispersion $-E_h(\mathbf{k})$ in the Moir\'e Brillouin zone for (a) the TLG-hBN $C = 3$ valence band, (b) TLG-hBN $C = 0$ valence band, (c) TBG-hBN $C = 1$ valence band with $w_1 = 110$ meV, and (d) TBG-hBN $C = 1$ valence band with $w_1 = 90$ meV. $w_1$ is the off-diagonal tunneling strength in TBG~\cite{supple}. The inverse screening length is $\kappa = 1/a_M$. } 
	\label{fig1}
\end{figure}

{\em Particle-hole duality and emergent Fermi liquids.}
From now on, we drop the spin and valley indices for simplicity. Upon a particle-hole transformation $ c(\mathbf{k}) \rightarrow d^{\dagger}(\mathbf{k}) $, the interaction Hamiltonian \eqref{onesector} becomes 
\begin{multline} 
H^{\rm proj} \rightarrow \sum_{\mathbf{k} \in \text{MBZ}} E_h(\mathbf{k}) d^\dagger(\mathbf{k})d(\mathbf{k}) +  \\ \sum_{\{\mathbf{k_i}\}\in \text{MBZ}}V^*_{\mathbf{k}_1\mathbf{k}_2\mathbf{k}_3\mathbf{k}_4} d^\dagger(\mathbf{k}_1) d^\dagger(\mathbf{k}_2) d(\mathbf{k}_3) d(\mathbf{k}_4) 
\end{multline} 
with \begin{equation}
	E_{h}(\mathbf{k}) = \sum_{\mathbf{k}'\in \text{MBZ}} \big( V_{\mathbf{k}'\mathbf{k}\mathbf{k}'\mathbf{k}} +  V_{\mathbf{k}\mathbf{k}'\mathbf{k}\mathbf{k}'} - V_{\mathbf{k}\mathbf{k}'\mathbf{k}'\mathbf{k}}-V_{\mathbf{k}'\mathbf{k}\mathbf{k}\mathbf{k}'}\big).\label{Eh}
\end{equation}
As a result, the dual Hamiltonian expressed in terms of hole operators acquires an interaction-induced single-hole dispersion which is nonconstant for generic projected interactions, as shown in Fig.~\ref{fig1}. For some simple toy flatband models designed for finding fractional Chern insulators in systems with a small unit cell, it has been observed that such a dispersion dominates over the hole-hole interaction at large electron fillings $\nu\gtrsim 4/5$, thus destabilizing FCIs and resulting in Fermi liquidlike states~\cite{lauchli2013hierarchy,Emilreview}.

\begin{figure*}
	\includegraphics[width=2.14\columnwidth]{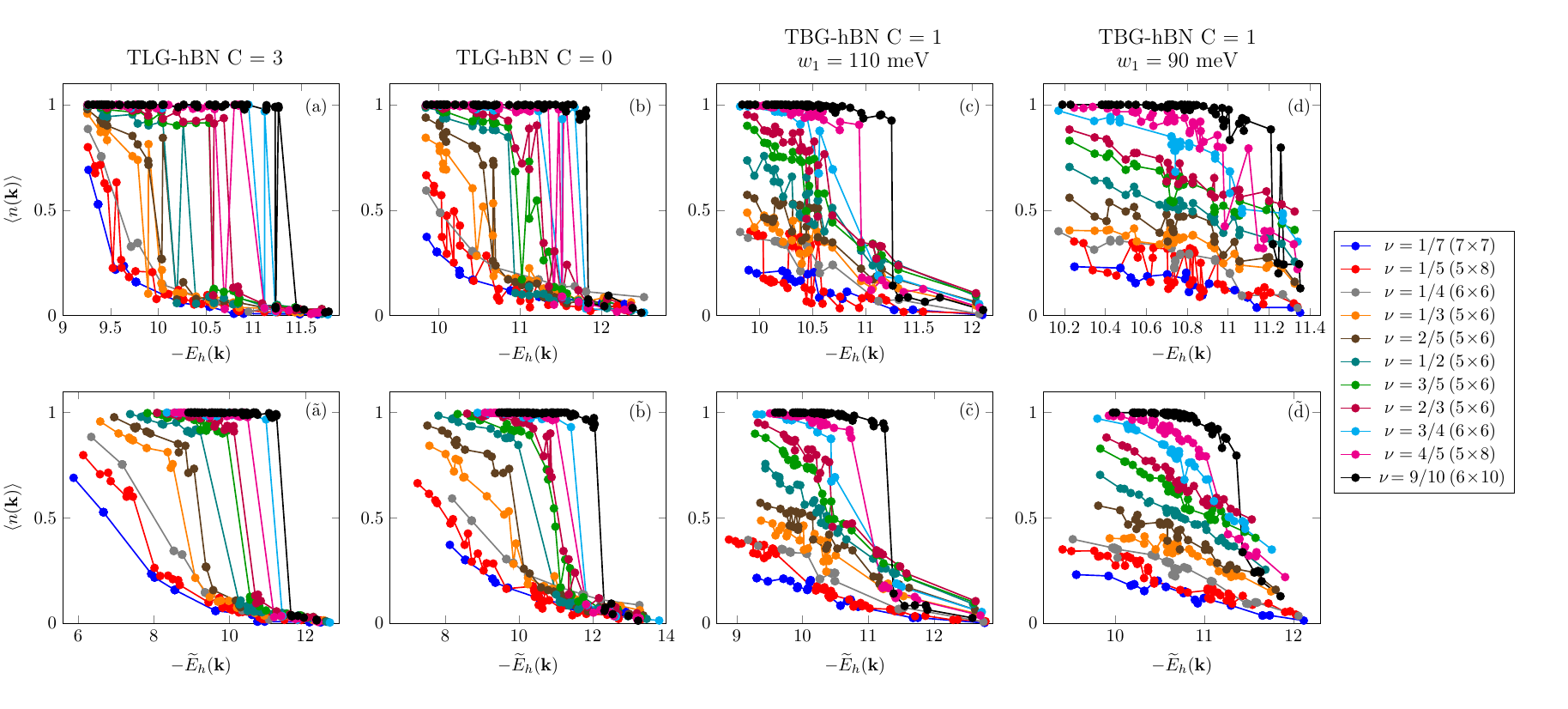}
	\caption{ The ground-state electron density $\langle n(\mathbf{k}) \rangle$ versus [(a)-(d)] the original hole dispersion $-E_h(\mathbf{k})$ and [($\tilde{\rm a}$)-($\tilde{\rm d}$)] the renormalized hole dispersion $-\widetilde{E}_h(\mathbf{k})$ at various electron fillings $\nu$. Note the different off-diagonal tunnelling strength $w_1$ between two graphene layers~\cite{supple} in [(c),($\tilde{\rm c}$)] and [(d),($\tilde{\rm d}$)]. The parentheses next to $\nu$ denote the system size $N_1\times N_2$. The inverse screening length is $\kappa = 1/a_M$. } 
	\label{fig2}
	\end{figure*}

Remarkably, we find that the effect of the particle-hole symmetry breaking terms [Eq.~(\ref{Eh})] is much more pronounced and dictates the physics also far away from the dilute holelimit in the experimentally relevant Moir\'e flatband systems. This is manifest in Figs.~\ref{fig2}(a)-(d) that display the ground-state electron density $\langle n(\mathbf{k}) \rangle = \langle c^\dagger(\mathbf{k}) c (\mathbf{k}) \rangle $ parametrized by the effective single-hole energy $-E_h(\mathbf{k})$ at various electron fillings $1/7\leq\nu\leq9/10$ in TLG-hBN and TBG-hBN. In all cases that we study, the electron density is quite generally strongly dependent on -- and essentially a monotonic function of -- the effective single-hole energy. This correspondence becomes even more striking when Hartree-Fock corrections to the single-hole energy are accounted for with a renormalized hole dispersion~\cite{supple}
\begin{equation} \widetilde{E}_h(\mathbf{k})\! =\! \frac{1}{\nu} \sum_{\mathbf{k}'} \langle n(\mathbf{k}') \rangle( V_{\mathbf{k}\mathbf{k}'\mathbf{k}\mathbf{k}'} +  V_{\mathbf{k}'\mathbf{k}\mathbf{k}'\mathbf{k}} - V_{\mathbf{k}\mathbf{k}'\mathbf{k}'\mathbf{k}}-V_{\mathbf{k}'\mathbf{k}\mathbf{k}\mathbf{k}'}),  \end{equation} 
which leads to a slightly modified parametrization of  $\langle n(\mathbf{k}) \rangle$ as shown in Figs.~\ref{fig2}($\tilde{\rm a}$)-($\tilde{\rm d}$). 

Saliently, we find clearly resolved emergent Fermi liquid behavior signaled by a sharp step in the electronic occupation numbers at electron fillings
extending all the way down to $ \nu = 1/3$ for the valence bands of TLG-hBN. While this is already quite clear from the bare single-hole dispersion [Figs.~\ref{fig2}(a)-(b)], the renormalized hole energy removes occasional oscillations of the density close to the emergent Fermi surface, making it remarkably well defined despite the relatively small systems available in our exact diagonalization study [Figs.~\ref{fig2}($\tilde{\rm a}$)-($\tilde{\rm b}$)]. The emergent Fermi surfacelike structure naturally becomes increasingly sharp at increasing $\nu$, corroborating that the ground states at those fillings are indeed weakly interacting Fermi liquids parametrized by the hole dispersion. Although the Fermi surface feature starts to dissolve below $\nu = 1/3$, the strong correlation between electron density and hole dispersion persists even at fillings as low as $\nu=1/7$ [Figs.~\ref{fig2}(a)-(b) and ($\tilde{\rm a}$)-($\tilde{\rm b}$)], suggesting that the hole dispersion still plays an important role in the low-energy physics at very low electron fillings and that perturbative approaches can be applied to analyze at least some of the resulting instabilities.

Albeit still significant, the correspondence between the electron density and the emergent hole dispersion is less pronounced for the TBG-hBN valence band with $ C = 1 $. In this case, we generally observe more uniform electron densities without clear Fermi surface features at small electron fillings and emergent Fermi surfaces only at comparably larger $\nu$ [Figs.~\ref{fig2}(c) and ($\tilde{\rm c}$)]. At slightly weaker tunneling between two graphene layers this tendency is further enhanced [Figs.~\ref{fig2}(d) and ($\tilde{\rm d}$)] as can be rationalized from the reduced bandwidth of the single-hole dispersion [compare Fig.~\ref{fig1}(d) with Figs.~\ref{fig1}(a)-(c)]. TBG-hBN thus stands out as a particularly promising host for genuinely strongly correlated topological states like fractional Chern insulators.

\begin{figure*}
	\includegraphics[width=2.1\columnwidth]{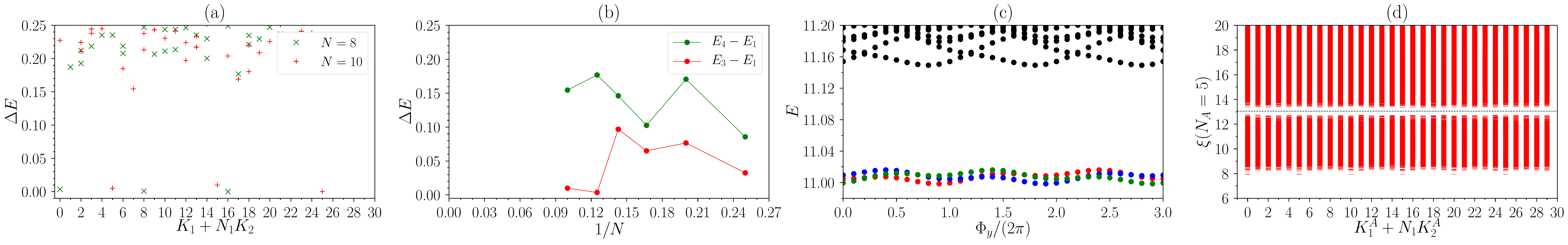}
	\caption{Evidence of $\nu=1/3$ FCIs in the $C=1$ band of TBG-hBN. Here we choose $w_1=90{\rm meV}$~\cite{supple} and the inverse screening length is $\kappa = 1/a_M$. (a) The low-lying energy spectrum for $N=8, N_1\times N_2=4\times6$ and $N=10,N_1\times N_2=5\times6$. (b) The finite-size scaling of the energy gap (green) and the ground-state splitting (red) for $N=4-10$. We define the energy gap and the ground-state splitting as $E_4-E_1$ and $E_3-E_1$, respectively, where $E_i$ is the $i$th energy level in ascending order. (c) The spectral flow for $N=10,N_1\times N_2=5\times 6$, where $\Phi_y$ is the magnetic flux insertion in the ${\bf a}_2$-direction. (d) The particle entanglement spectrum for $N=10,N_1\times N_2=5\times 6$, with $23256$ levels below the entanglement gap (the dashed line).}
	\label{fig3}
\end{figure*}

{\em Fractional Chern insulators.-} 
Now we address the possibility of FCIs stabilized by the screened Coulomb interaction in the $C=1$ band of TBG-hBN. 
We consider electronic band filling $\nu = 1/3$ where one may on general grounds anticipate that FCIs may exist and focus the case with off-diagonal tunneling strength $w_1=90 \> {\rm meV}$ encouraged by the relatively weak single-hole dispersion [Fig.~\ref{fig1}(d)] and comparably featureless electron density profile [Figs.~\ref{fig2}(d) and ($\tilde{\rm d}$)] at this parameter value. As detailed below and in Fig.~\ref{fig3} we indeed find compelling numerical evidence of stable high-temperature $\nu=1/3$ FCIs.

First we examine the energetics of the Hamiltonian (\ref{onesector}). For various system sizes that we study, there are three approximately degenerate states at the bottom of the energy spectrum, whose total momentum $(K_1,K_2)$ are consistent with the prediction of Haldane statistics~\cite{fciprx,manybodysymmetry,haldanestatistics} for the $\nu=1/3$ FCIs. These three ground states are separated from excited states by a clear energy gap, which becomes larger than the ground-state splitting as the system size grows [Fig.~\ref{fig3}(a)]. A finite-size scaling of the energy gap demonstrates that it is very likely to survive in the thermodynamic limit [Fig.~\ref{fig3}(b)]. Moreover, the three fold ground-state degeneracy persists under magnetic flux insertion through the handles of the toroidal system [Fig.~\ref{fig3}(c)], which confirms the robustness of topological degeneracy.

To further corroborate that the ground states are topologically nontrivial, we have investigated the particle entanglement spectrum (PES), which contains information of the excitation structure of the system and can be used to distinguish FCIs from competing possibilities such as charge density waves~\cite{fciprx,LiH,PES}. The result is displayed in Fig.~\ref{fig3}(d), where we divide the whole system into $N_A$ and $N-N_A$ electrons and label each PES level by the total momentum $(K_1^A,K_2^A)$ of those $N_A$ electrons. We find a clear entanglement gap separating the low-lying PES levels from higher ones, and the number of levels below the gap exactly matches the pertinent counting of quasihole excitations in the $\nu=1/3$ Abelian FCI states~\cite{fciprx,manybodysymmetry,haldanestatistics}.

We have also considered the situation with larger interlayer tunneling in TBG-hBN. We increase $w_1$ to $110 \> {\rm meV}$, and again we observe clear signatures of FCI states albeit only at somewhat stronger screening $\kappa = 6/a_M$~\cite{supple}. Moreover, we find that projecting the interaction to the other $C=-1$ band of TBG-hBN gives very similar results to Fig.~\ref{fig3}~\cite{supple}, indicating that FCI states can exist in both topological bands of this model.

{\em Discussion.-}
We have shown that the dual description of Coulomb interactions in partially filled Moir\'e flatbands in terms of holes provides remarkable insights about this strongly correlated problem, turning it into an effectively weakly interacting problem in a large portion of the phase diagram. In particular, this identifies ABC stacked trilayer and twisted bilayer graphene aligned with boron nitride as realistic hosts of a new paradigm for emergent Fermi liquids deriving from essentially pure interaction effects at appropriate doping. This perspective opens up a number of intriguing experimental prospects as well as providing a key quantity -- the single-hole dispersion, $E_h(\bf k)$ -- with a bandwidth estimate that is typically significantly larger than the flatband bandwidth \cite{supple}. It enables the use of standard many-body techniques which are typically not applicable to flatband systems due to the apparent lack of a small parameter. In particular, our simple Hartree-Fock calculations for the holes corroborate the generic applicability of the interaction driven Fermi liquid phenomenology. 

The exotic interaction induced Fermi surfaces come with telltale experimental signatures in terms of quantum oscillations (cf. Fig. \ref{fig1}). In this context it is worth noting that there are plenty of experimentally tunable knobs that can alter the shape of these Fermi surfaces -- and eventually cause their breakdown. While some instabilities invite future research, such as those leading to a convex density profile at low filling fractions in the trilayer graphene setup (cf. Fig. \ref{fig2}), we have investigated in detail the case when the Fermi surface is smeared in the most extreme sense with an essentially featureless density profile only weakly linearly correlated with the hole energy. Here we provided numerical evidence that fractional Chern insulators may form in TBG-hBN at electronic band filling $\nu=1/3$. It is quite remarkable that such state is stabilized in a realistic model systems with long-range Coulomb interactions (screening length $\sim a_M\sim 60$ graphene lattice constants \cite{abinitiomoire}) in stark contrast to previously studied toy models which generically feature strictly short-range interactions. Here, again, the hole dispersion provided key intuition: only for models with a small hole bandwidth there is a chance of realizing FCIs. This in turn allows us to develop an intuition for which externally tunable parameters that may be alter the phase diagram to achieve the desired many-body states. In particular, we have found either changing the screening length or the interlayer tunneling, which are tunable through pressure and substrate engineering respectively, can effectively (de)stabilize FCIs~\cite{supple}. 

Another remarkable feature of the interacting Moir\'e flatbands is the large energy scale associated with the projected Coulomb interaction; a rough estimate of the stability of the FCIs based on the finite-size gap calculations (cf. Fig. \ref{fig3}) suggests that they may persist all the way up to $\sim 10$ K. This thus identifies twisted bilayer graphene aligned with boron nitride as a particularly promising platform for finding observing the first bona fide FCIs -- at elevated temperatures and in absence of an external magnetic field. Moreover, our numerical evidence of $\nu=1/3$ FCIs in TBG-hBN, invites a systematic search for fractionalized topological states at other fillings and in Moir\'e flatbands derived from other materials.

Finally we stress that we have thus far only considered a single fractionally filled spin-valley sector. While this is a very reasonable assumption for spatially homogenous phases, the multicomponent nature of the problem and effects of symmetry breaking are expected to further enrich the phenomenology of interacting Moir\'e flatbands. Also in this context we suggest that the dual formulations in terms of holes will likely provide key physical insights.

\acknowledgments
{\it Acknowledgments.-}
We would like to thank Johan Carlstr\"om and Roderich Moessner for useful discussions. A.A. and E.J.B. are supported by the Swedish Research Council (VR) and the Wallenberg Academy Fellows program of the Knut and Alice Wallenberg Foundation. Z.L. is supported by the National Natural Science Foundation of China through Grant No. 11974014.

\newpage
\onecolumngrid 
\newpage
\def\theequation{S\arabic{equation}}
\section{Supplementary Material}
In this Supplementary Material we provide technical details of the single-particle Hamiltonians of TLG-hBN and TBG-hBN, the projected interaction, the Hartree-Fock corrections to the single-hole energy, a comparison between the single-hole dispersion and the flat valence band dispersion and further numerics corroborating the conclusions about fractional Chern insulators in TBG-hBN.
\subsection{TLG-hBN single-particle model}
First we consider an ABC-stacked trilayer graphene whose lattice constant is $a\approx 2.46{\rm \AA}$. At low energy, the physics around one of the valleys can be effectively described in terms of the sublattice degrees of freedom $A_1$ in the top layer and $B_3$ in the bottom layer, where $A$ and $B$ correspond to the two sublattices in a single-layer graphene. Aligning the top layer of the trilayer graphene with hexagonal Boron Nitride (hBN) induces a Moir\'e potential which results in a superlattice structure with lattice constant $a_M \approx 60 \> a$ \cite{abinitiomoire}. We assume that there is no twist angle between hBN and the top graphene layer.

We focus on the $K$ valley, and use the same single-particle Hamiltonian of TLG-hBN for each spin favor as in Ref.~\cite{PhysRevLett.122.016401}: 
\begin{equation}
H = \sum_{\mathbf{k}} \psi^{\dagger} (\mathbf{k})\mathcal{H}_0(\mathbf{k}) \psi(\mathbf{k})  + \sum_{\mathbf{k}}\sum_{i = 1}^6 \psi^{\dagger}(\mathbf{k}+\mathbf{G}_i) H_M(\mathbf{G}_i) \psi(\mathbf{k}) 
\label{eqs6}
\end{equation}
with $\psi(\mathbf{k}) = \begin{pmatrix}
\psi_{A_1}(\mathbf{k})\\ \psi_{B_3}(\mathbf{k})
\end{pmatrix} $. 
In the first term of $H$, we have $ \mathcal{H}_0(\mathbf{k}) = \frac{(\hbar v_0)^3}{(t_1)^2} \begin{pmatrix}
0 &  (k_x - ik_y)^3\\ 
(k_x + ik_y)^3 & 0
\end{pmatrix}  + U \sigma_z + H_R $, where $U$ is an applied voltage along the trilayer graphene, $H_R$ describes remote hopping corrections to the model \cite{PhysRevB.82.035409}, and we use the same hopping parameters as in Ref.~\cite{PhysRevLett.122.016401}. In the second term of $H$, the Moir\'e hoppings $H_M$ which act only on the top graphene layer are given by 
\begin{equation}
H_M(\mathbf{G}_1) = H_M(\mathbf{G}_3) = H_M(\mathbf{G}_5) = \begin{pmatrix}
C_0 e^{i\phi_0} + C_z e^{i\phi_z} & 0 \\ 
0 & 0
\end{pmatrix} 
\end{equation}
and 
\begin{equation}
H_M(\mathbf{G}_2) = H_M(\mathbf{G}_4) = H_M(\mathbf{G}_6) = \begin{pmatrix}
C_0 e^{-i\phi_0} + C_z e^{-i\phi_z} & 0 \\ 
0 & 0
\end{pmatrix},  
\end{equation}
where $\mathbf{G}_1 = (0,\frac{4\pi}{\sqrt{3} a_M }) $ and the rest of $\mathbf{G}_i $ are obtained by rotating $\mathbf{G}_1$ by sixty degrees successively, $C_0 = -10.13 \> \text{meV}$, $C_z = -9.01 \> \text{meV}$, $\phi_0 = 86.53^\circ $, and $\phi_z = 8.43^\circ$. 

For each ${\bf k}_0$ in the Moir\'e Brillouin zone (MBZ), by writing ${\bf k}$ in the single-particle Hamiltonian $H$ as ${\bf k}_0+m{\bf G}_1+n{\bf G}_2$ and setting integers $m,n=-d,...,d$, we can construct $H$ as a matrix of dimension $2(2d+1)^2$. The eigenvalues and eigenvectors of this Hamiltonian matrix then give us the band structure of TLG-hBN. We use $U = 10$ meV in our numerics, which results in a nearly flat valence band with Chern number $ C = 3 $. Switching the sign of $U$ changes the Chern number to $C = 0$.

\subsection{TBG-hBN single-particle model}
We now consider a twisted bilayer graphene (TBG) with twist angle $\theta$.
At a set of magic angles, the system has very flat valence and conduction bands around charge neutrality \cite{twistedbilayermodel1,twistedbilayermodel2,twistedbilayermodel3}. Aligning the system with hBN would break the $C_2 \mathcal{T}$ symmetry and hence gaps out all the protected crossings \cite{originofhbngap1,originofhbngap2,originofhbngap3,originofhbngap4}. The alignment with hBN would induce a staggered onsite potential on the top graphene layer. Moreover, it will induce another Moir\'e potential which is incommensurate with but much weaker than the original Moir\'e potential of TBG. To the lowest order of approximation, the Moir\'e potential induced by hBN can be neglected, so we only keep the onsite potential to simulate the effect of hBN here. 

We focus on the Moir\'e Brillouin zone of TBG near the valley $\mathbf{K}_{+} = \frac{4\pi}{3 a}(1,0) $ of single-layer graphene, where $a\approx 2.46{\rm \AA}$ is the lattice constant of graphene. The two primitive reciprocal lattice vectors of TBG are chosen as ${\bf G}_1=\frac{4\pi}{a}\sin\frac{\theta}{2}(\frac{1}{\sqrt{3}},1)$ and ${\bf G}_1=\frac{4\pi}{a}\sin\frac{\theta}{2}(-\frac{1}{\sqrt{3}},1)$. Let us denote ${\bf K}_+^t=R_{\theta/2}{\bf K}_+$ and ${\bf K}_+^b=R_{-\theta/2}{\bf K}_+$, where $R_{\theta} $ a counter-clockwise rotation around the $z$-axis in the momentum space. The single-particle Hamiltonian of TBG-hBN for each spin favor can then be written as 
\begin{eqnarray}
H &=&  \sum_{\mathbf{k}} \psi^{\dagger}_t(\mathbf{k}) h_{-\theta/2}(\mathbf{k}-{\bf K}_+^t) \psi_{t}(\mathbf{k}) + \sum_{\mathbf{k}} \psi^{\dagger}_b(\mathbf{k}) h_{\theta/2}(\mathbf{k}-{\bf K}_+^b) \psi_{b}(\mathbf{k}) \nonumber\\
&+& \sum_{\mathbf{k}}\sum_{ j = 0}^2 \big(\psi_{t}^{\dagger}(\mathbf{k}-{\bf q}_0+\mathbf{q}_j) T_j \psi_{b}(\mathbf{k}) + h.c. \big)  + M \sum_{\mathbf{k}} \psi_{t}^{\dagger}(\mathbf{k}) \sigma_z \psi_{t}(\mathbf{k}) ,
\label{eqs9}
\end{eqnarray}
where $ \psi_t({\bf k}) = \begin{pmatrix}
\psi_{A_t}(\mathbf{k})\\ \psi_{B_t}(\mathbf{k})
\end{pmatrix}$ and $\psi_{b}({\bf k})= \begin{pmatrix}
\psi_{A_b}(\mathbf{k})\\ \psi_{B_b}(\mathbf{k})
\end{pmatrix}$ are spinors of annihilation operators for electrons in top and bottom graphene layers, respectively, and $A$ and $B$ correspond to the two sublattices in single-layer graphene. In the first two terms of $H$,  we define $h_{\theta}(\mathbf{k}) = h(R_{\theta} \mathbf{k}) $ where $h(\mathbf{k}) = -\frac{\sqrt{3}}{2} a t_0 (k_x \sigma_x + k_y \sigma_y)$ with $t_0\approx 2.62{\rm eV}$ is the standard Dirac Hamiltonian of single-layer graphene. The last term is the onsite potential induced by hBN on the top graphene layer. The third term, i.e., the Moir\'e hoppings in twisted bilayer graphene are given by
\begin{equation}
T_j = w_0 - w_1 e^{i(2\pi/3)j \sigma_z} \sigma_x e^{-i(2\pi/3)j\sigma_z}
\end{equation}
and $\mathbf{q}_0 = R_{-\theta/2} \mathbf{K}_+ - R_{\theta/2} \mathbf{K}_+$, $\mathbf{q}_1  = R_{2\pi/3} \mathbf{q}_0$ and $\mathbf{q}_2 = R_{-2\pi/3} \mathbf{q}_0$. 

For each ${\bf k}_0$ in the MBZ, by writing ${\bf k}$ in the single-particle Hamiltonian $H$ as ${\bf k}_0+m{\bf G}_1+n{\bf G}_2$ and setting integers $m,n=-d,...,d$, we can construct $H$ as a matrix of dimension $4(2d+1)^2$. The eigenvalues and eigenvectors of this Hamiltonian matrix then give us the band structure of TBG-hBN.  In our numerics, we take $ \theta = 1.05^\circ $, $ M = 15$ meV, $ w_1 = 110 $ meV and $90$ meV, and $w_0 = 0.7 \> w_1$. With these parameters, one can obtain two gapped flat valence and conduction bands around charge neutrality with Chern numbers $ C = 1 $ and $ C = -1 $, respectively. 

\subsection{Interaction Hamiltonian}
The density operator in valley $a$ with spin $\sigma$ can be written as
\begin{equation}
\rho_{a\sigma}(\mathbf{q}) = \sum_{\mathbf{k}}\sum_{\alpha} f^\dagger_{a\sigma;\alpha}(\mathbf{k} + \mathbf{q}) f_{a\sigma;\alpha}(\mathbf{k}).
\end{equation} 
Here $f_{a\sigma;\alpha}(\mathbf{k})$ is the microscopic electron operator and $\alpha$ is an orbital index which can take the layer ($t,b$) or the sublattice ($A,B$) depending on the model of interest. By writing $ \mathbf{k} = \mathbf{k}_1 + m_1 \mathbf{G}_1 + n_1 \mathbf{G}_2$ and $ \mathbf{k} + \mathbf{q} = \mathbf{k}_2 + m_2 \mathbf{G}_1 + n_2 \mathbf{G}_2$, where $\mathbf{k}_1$, $\mathbf{k}_2$ are restricted in the MBZ, $\mathbf{G}_1,\mathbf{G}_2$ are the primitive Moir\'e reciprocal lattice vectors and $m,n$ are integers, we have
\begin{equation}
\rho_{a\sigma}(\mathbf{q}) = \sum_{\mathbf{k}_1,\mathbf{k}_2 \in \text{MBZ}} \sum_{\alpha} \sum_{\{m_i,n_i\} = - d}^{d} \delta_{\mathbf{k}_2 - \mathbf{k}_1 + (m_2-m_1)\mathbf{G}_1 + (n_2 - n_1)\mathbf{G}_2,\mathbf{q}} \> f_{a\sigma;\alpha}^\dagger(\mathbf{k}_2 + m_2 \mathbf{G}_1 + n_2 \mathbf{G}_2) f_{a\sigma;\alpha}(\mathbf{k}_1 + m_1 \mathbf{G}_1 + n_1 \mathbf{G}_2),
\end{equation}
where $d$ is an appropriately chosen cutoff (we use $d=5$ in numerics). Because 
\begin{equation}
f_{a\sigma;\alpha}(\mathbf{k} + m\mathbf{G}_1 + n\mathbf{G}_2 ) = \sum_s \mu_{m,n,\alpha}^{s,a\sigma}(\mathbf{k}) c^s_{a\sigma}({\mathbf{k}}),
\end{equation} 
where $ \{\mu_{m,n,\alpha}^{s,a\sigma}(\mathbf{k}) \}$ is the eigenvector of the $s$th band in valley $a$ with spin $\sigma$ obtained from diagonalizing the single-particle Hamiltonian matrix and $c_{a\sigma}^s({\mathbf{k}})$ is the annihilation operator of an electron in that band, we obtain the density operator projected to a specific band in valley $a$ with spin $\sigma$ as
\begin{equation}
\tilde{\rho}_{a\sigma}(\mathbf{q}) = \sum_{\mathbf{k}_1,\mathbf{k}_2 \in \text{MBZ}} \sum_{\alpha} \sum_{\{m_i,n_i\} = - d}^{d} \delta_{\mathbf{k}_2 - \mathbf{k}_1 + (m_2-m_1)\mathbf{G}_1 + (n_2 - n_1)\mathbf{G}_2,\mathbf{q}} \>  \mu_{m,n,\alpha}^{*a\sigma}(\mathbf{k}_2)\mu_{m,n,\alpha}^{a\sigma}(\mathbf{k}_1) c_{a\sigma}^\dagger({\mathbf{k}}_2) c_{a\sigma}({\mathbf{k}}_1),
\label{eq14}
\end{equation}
where we have dropped the band index $s$.

The normal ordered Hamiltonian of a generic long-range interaction takes the form of 
\begin{equation}
\label{eq:int}
H = \frac{1}{2}\sum_{\mathbf{q} \in R^2 } \sum_{a_1 \sigma_1}\sum_{ a_2 \sigma_2} V(\mathbf{q}) : \rho_{a_1\sigma_1}(\mathbf{q}) \rho_{a_2,\sigma_2}(-\mathbf{q}) : ,
\end{equation} 
where $V({\bf q})$ is the Fourier transform of the interaction potential. For the screened Coulomb interaction, we have $V({\bf q})=\frac{e^2}{4\pi\epsilon S}\frac{2\pi}{\sqrt{|{\bf q}|^2+\kappa^2}}$, where $e$ is the electron charge, $\epsilon$ is the dielectric constant of the material, $S$ is the area of the Moir\'e superlattice, and $\kappa$ measures the screening strength. Notice that for a long-range interaction $V(\mathbf{q})$ is not periodic under Moir\'e reciprocal lattice translations, i.e., $V(\mathbf{q} + m\mathbf{G}_1 + n\mathbf{G}_2) \neq V(\mathbf{q}) $, so we need to sum over $ R^2$ in Eq.~(\ref{eq:int}). By substituting Eq.~(\ref{eq14}) into Eq.~(\ref{eq:int}), we end up with the projected Hamiltonian:
\begin{multline}
H^{\text{proj}} = \frac{1}{2} \sum_{\mathbf{q} \in R^2 } \sum_{a_1 \sigma_1 }\sum_{a_2 \sigma_2} V(\mathbf{q}) :\!\tilde{\rho}_{a_1\sigma_1}(\mathbf{q}) \tilde{\rho}_{a_2\sigma_2}(-\mathbf{q}) \!: \>  = \frac{1}{2}\sum_{\{\mathbf{k}_i\} \in \text{MBZ}} \sum_{\mathbf{q} \in R^2} \sum_{a_1 \sigma_1}\sum_{ a_2 \sigma_2} \sum_{\alpha \beta} \sum_{\{m_i,n_i\} = -d}^{d} \\ \delta_{\mathbf{k}_1 + \mathbf{k}_2 + (m_1+m_2)\mathbf{G}_1 + (n_1+n_2)\mathbf{G}_2, \mathbf{k}_3 + \mathbf{k}_4 + (m_3+m_4)\mathbf{G}_1 + (n_3+n_4)\mathbf{G}_2} 
\delta_{\mathbf{k}_1 - \mathbf{k}_4 + (m_1-m_4)\mathbf{G}_1 + (n_1 - n_4)\mathbf{G}_2,\mathbf{q}}  \\ \times V(\mathbf{q}) \mu^{*a_1\sigma_1}_{m_1,n_1,\alpha}(\mathbf{k}_1) \mu^{*a_2 \sigma_2}_{m_2,n_2,\beta}(\mathbf{k}_2) \mu^{a_2\sigma_2}_{m_3,n_3,\beta}(\mathbf{k}_3)\mu^{a_1\sigma_1}_{m_4,n_4,\alpha}(\mathbf{k}_4) c_{a_1\sigma_1}^\dagger({\mathbf{k}}_1) c_{a_2\sigma_2}^\dagger({\mathbf{k}}_2) c_{a_2\sigma_2}({\mathbf{k}}_3) c_{a_1\sigma_1}({\mathbf{k}}_4),
\label{eqs16}
\end{multline}
which can be further simplified to Eq.~(2) in the main text under the assumption of polarized spin and valley degrees of freedom. The $\delta$ function in Eq.~(\ref{eqs16}) guarantees the crystal momentum conservation in the projected interaction. However, because we use the effective single-particle models Eqs.~(\ref{eqs6}) and (\ref{eqs9}) to calculate $ \mu_{m,n,\alpha}^{a\sigma}(\mathbf{k}) $, the usual identity between the many-body energy levels in the $(K_1,K_2)$ and $(N_1-K_1,N_2-K_2)$ sectors does not exactly hold for a finite lattice with $N_1\times N_2$ unit cells.

\subsection{Renormalized hole energy} 
Starting from the interaction Hamiltonian for holes, \begin{equation} H^{\text{proj}} = 
\sum_{\mathbf{k} \in \text{MBZ}} E_h(\mathbf{k}) d_{\mathbf{k}}^\dagger d_{\mathbf{k}} + \sum_{\substack{\mathbf{k}_1\mathbf{k}_2\mathbf{k}_3\mathbf{k}_4\\\in \text{MBZ}}}V_{\mathbf{k}_1\mathbf{k}_2\mathbf{k}_3\mathbf{k}_4}^{*} d_{\mathbf{k}_1}^\dagger d_{\mathbf{k}_2}^\dagger d_{\mathbf{k}_3} d_{\mathbf{k}_4}
\end{equation}
we can perform a simple mean-field approximation to the two-body interaction. We first replace the two-body interaction terms with the Hartree and Fock terms \begin{equation}
H_{\text{Hartree}} = \sum_{\substack{\mathbf{k}_1\mathbf{k}_2\mathbf{k}_3\mathbf{k}_4\\\in \text{MBZ}}} V_{\mathbf{k}_1\mathbf{k}_2\mathbf{k}_3\mathbf{k}_4}^{*} \bigg(\langle d_{\mathbf{k}_1}^\dagger d_{\mathbf{k}_4}\rangle d_{\mathbf{k}_2}^\dagger d_{\mathbf{k}_3} + \langle d_{\mathbf{k}_2}^\dagger d_{\mathbf{k}_3}\rangle d_{\mathbf{k}_1}^\dagger d_{\mathbf{k}_4} - \langle d_{\mathbf{k}_1}^\dagger d_{\mathbf{k}_4}\rangle \langle d_{\mathbf{k}_2}^\dagger d_{\mathbf{k}_3}\rangle\bigg) 
\end{equation} 
and 
\begin{equation}
H_{\text{Fock}} = \sum_{\substack{\mathbf{k}_1\mathbf{k}_2\mathbf{k}_3\mathbf{k}_4\\\in \text{MBZ}}} V_{\mathbf{k}_1\mathbf{k}_2\mathbf{k}_3\mathbf{k}_4}^{*} \bigg(-\langle d_{\mathbf{k}_1}^\dagger d_{\mathbf{k}_3}\rangle d_{\mathbf{k}_2}^\dagger d_{\mathbf{k}_4} - \langle d_{\mathbf{k}_2}^\dagger d_{\mathbf{k}_4}\rangle d_{\mathbf{k}_1}^\dagger d_{\mathbf{k}_3} + \langle d_{\mathbf{k}_1}^\dagger d_{\mathbf{k}_3}\rangle \langle d_{\mathbf{k}_2}^\dagger d_{\mathbf{k}_4}\rangle\bigg).
\end{equation}
Then by assuming $\langle d_{\mathbf{k}_1}^\dagger d_{\mathbf{k}_4}\rangle = \delta_{\mathbf{k}_1,\mathbf{k}_4} \langle d_{\mathbf{k}_1}^\dagger d_{\mathbf{k}_1}\rangle $ and making use of $\mathbf{k}_1 + \mathbf{k}_2 = \mathbf{k}_3 + \mathbf{k}_4 $, we obtain the renormalized single-hole dispersion as \begin{equation}
\widetilde{E}_h(\mathbf{k})\! =\! \frac{1}{\nu} \sum_{\mathbf{k}'} \langle n(\mathbf{k}') \rangle( V_{\mathbf{k}\mathbf{k}'\mathbf{k}\mathbf{k}'} +  V_{\mathbf{k}'\mathbf{k}\mathbf{k}'\mathbf{k}} - V_{\mathbf{k}\mathbf{k}'\mathbf{k}'\mathbf{k}}-V_{\mathbf{k}'\mathbf{k}\mathbf{k}\mathbf{k}'}),
\end{equation} 
where we divide by the filling factor $\nu$ so that the range of the renormalized single-hole dispersion is similar at all fillings.
\begin{figure}[t!]
	\includegraphics[width=\columnwidth]{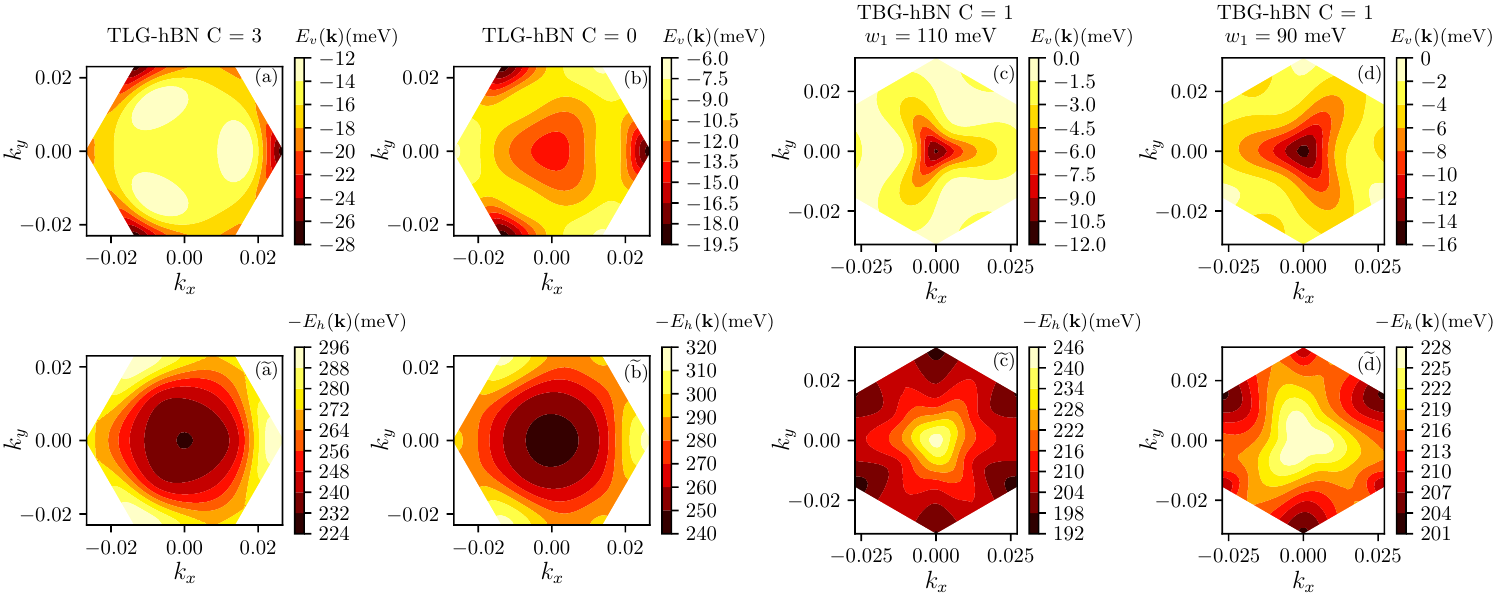}
	\caption{Contour plots for [(a)-(d)] The flat valence band dispersion $E_v(\mathbf{k})$ and [($\tilde{\rm a}$)-($\tilde{\rm d}$)] the single-hole dispersion $-E_h(\mathbf{k})$. Notice the bandwidth of each dispersion as indicated by the range of the energy values in the color bar besides each plot. The inverse screening length is $\kappa = 1/a_M$. }
	\label{bandwidth}
\end{figure}

\subsection{The single-hole dispersion vs Flatband dispersion}
In this section, we highlight the importance of the singlehole dispersion by comparing its bandwidth with the bandwidth of the original flatband dispersion for both TLG-hBN and TBG-hBN. For TLG-hBN, it has been suggested \cite{chen2019evidence,Chen2019} that the strength of the Coloumb interaction is about 25 meV. We find, as shown in Figs. \ref{bandwidth}($\tilde{\text{a}}$)-\ref{bandwidth}($\tilde{\text{b}}$), that this corresponds to a bandwidth of the single-hole dispersion that is about 70 meV and 80 meV for the C = 3 and C = 0 valence bands respectively. That is indeed much larger than the flatband bandwidth [Figs. \ref{bandwidth}(a)-\ref{bandwidth}(b)] that is approximately around 15 meV for both bands. On the other hand, the strength of the Coloumb interactions in TBG-hBN is estimated to be around 20 meV or possibly larger \cite{cao2018correlated,spectrobilayer}. This would in turn corresponds to a single-hole dispersion bandwidth [Figs. \ref{bandwidth}($\tilde{\text{c}}$)-\ref{bandwidth}($\tilde{\text{d}}$)] of about 55 meV and 27 meV compared to flatband bandwidth [Figs. \ref{bandwidth}(c)-\ref{bandwidth}(d)] of about 12 meV and 16 meV for $w_1 = 110$ meV and $w_1 = 90$ meV respectively. We note that as a matter of principle one can alter the strength of the interaction by changing the substrate. As this is something that can in principle be engineered and we wish to isolate the fundamental impact of interactions, we have in this work mainly ignored the band-dispersion. Although it in general will have a quantitative impact, it is for the experimentally relevant regime that we do not expect it to change the physics qualitatively.  

\subsection{More numerical results for the $\nu=1/3$ FCIs in TBG-hBN}
In this section, we present more numerical results to support the existence of stable $\nu=1/3$ FCIs in the topological flat bands of TBG-hBN. 

\begin{figure}
	\includegraphics[width=\columnwidth]{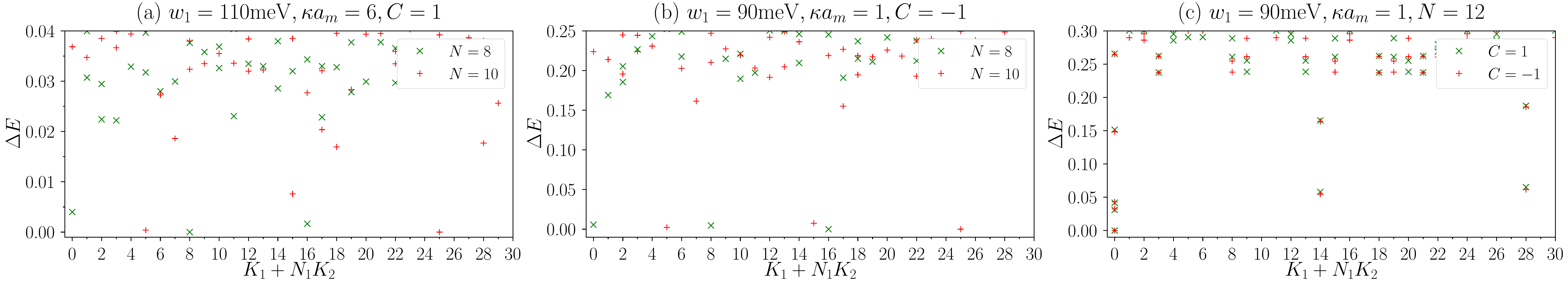}
	\caption{More numerical results for the $\nu=1/3$ FCIs in TBG-hBN. (a) The low-lying energy spectra of the screened Coulomb interaction projected to the $C=1$ band for $N=8, N_1\times N_2=4\times6$ and $N=10,N_1\times N_2=5\times6$ at a stronger interlayer tunnelling $w_1=110{\rm meV}$. The inverse screening length is $\kappa = 6/a_M$. (b) The low-lying energy spectra of the screened Coulomb interaction projected to the $C=-1$ band for $N=8, N_1\times N_2=4\times6$ and $N=10,N_1\times N_2=5\times6$ at $w_1=90{\rm meV}$. The inverse screening length is $\kappa = 1/a_M$. (c) The low-lying energy spectra of the screened Coulomb interaction projected to the $C=\pm1$ bands for $N=12, N_1\times N_2=6\times6$ at $w_1=90{\rm meV}$. The inverse screening length is $\kappa = 1/a_M$.}
	\label{fcimore}
\end{figure}

First, as stronger interlayer tunnellings were considered in some studies of twisted bilayer graphene~\cite{futwisted,2019arXiv190108209Z}, we increase $w_1$ from $90{\rm meV}$ to $110{\rm meV}$ and still keep $w_0=0.7w_1$ to study the many-body energy spectra of the screened Coulomb interaction projected to the $C=1$ flat band. At such a stronger tunnelling, according to our numerical results in small systems, we find that FCIs can emerge only if we enhance the Coulomb screening to $\kappa \approx 6/a_M$. As displayed in Fig.~\ref{fcimore}(a), we again observe clear three-fold ground-state degeneracies of FCI states at $\nu=1/3$ with $\kappa = 6/a_M$, although the energy spectra demonstrate stronger finite-size effects and larger ground-state splitting than those shown in the main text. In contrast, at weaker tunnelling $w_1=90{\rm meV}$ we can already see convincing FCI signatures with much smaller (even zero) screening. Our results suggest that the stability of FCIs in TBG-hBN requires stronger Coulomb screening as the two graphene layers are more coupled with each other. Hereafter we will stick with  $w_1=90{\rm meV}$ and $\kappa = 1/a_M$.

Second, we project the screened Coulomb interaction to the other topological flat band -- the $C=-1$ band in TBG-hBN above the charge neutrality. The obtained many-body spectra at electron filling $\nu=1/3$ [Fig.~\ref{fcimore}(b)] are very similar to those for the $C=1$ band. As the $C=-1$ band has a similar band width and an even larger band gap than the $C=1$ band, it is also a promising host of FCIs in TBG-hBN.

Finally, we would like to show the results of the largest system size, $N=12,N_1\times N_2=6\times 6$, which we can reach by exact diagonalization [Fig.~\ref{fcimore}(c)]. For this system size, we do find that the lowest three levels are located in the momentum sector $(K_1,K_2)=(0,0)$ as expected for $\nu=1/3$ FCIs, and these three states are quite close to degenerate in the $(K_1,K_2)=(0,0)$ sector. However, there are two other levels -- one in the $(K_1,K_2)=(2,2)$ sector and the other in the $(K_1,K_2)=(4,4)$ sector coming down quite close in in energy, thus the global three-fold ground-state degeneracy is no longer clearly separated from excited states. However, given the compelling results in other system sizes, we firmly believe that the absence of a clearly resolved FCI gap for $N=12,N_1\times N_2=6\times 6$ is a finite-size effect. Precisely such finite size effects are indeed known to be significantly enhanced for samples where the model ground states all resolve in the same momentum sector. 

\begin{thebibliography}{10}
\bibitem{cao2018unconventional}
Y. Cao, V. Fatemi, S. Fang, K. Watanabe, T. Taniguchi,
E. Kaxiras, and P. Jarillo-Herrero,
{\em Unconventional superconductivity in magic-angle graphene
	superlattices},
\href{https://www.nature.com/articles/nature26160}{Nature, {\bf 556}, 43 (2018)}.
\bibitem{cao2018correlated}
Y. Cao, V. Fatemi, A. Demir, S. Fang, S.L. Tomarken, J.Y. Luo, J.D. Sanchez-Yamagishi, K. Watanabe, T. Taniguchi, E. Kaxiras, et~al.
{\em Correlated insulator behaviour at half-filling in magic-angle
	graphene superlattices},
\href{https://www.nature.com/articles/nature26154}{Nature, {\bf{556}}, 80 (2018)}.
\bibitem{yankowitz2019tuning}
M. Yankowitz, S. Chen, H. Polshyn, Y. Zhang, K. Watanabe,
T. Taniguchi, D. Graf, A.F. Young, and C.R. Dean,
{\em Tuning superconductivity in twisted bilayer graphene,}
\href{https://science.sciencemag.org/content/363/6431/1059}{ Science, {\bf 363}, 1059 (2019)}.
\bibitem{Lu2019}
X. Lu, P. Stepanov, W. Yang, M. Xie, M.A. Aamir, I. Das,
C. Urgell, K. Watanabe, T. Taniguchi, G. Zhang, A. Bachtold, A.H. MacDonald, and D.K. Efetov,
{\em Superconductors, orbital magnets and correlated states in magic-angle
	bilayer graphene. } \href{https://www.nature.com/articles/s41586-019-1695-0}{Nature, {\bf 574}, 653 (2019).}
\bibitem{spectrobilayer}
Y. Xie, B. Lian, B. Jäck, X. Liu, C. Chiu, K. Watanabe. T. Taniguchi, B.A. Bernevig and A. Yazdani
{\em Spectroscopic signatures of many-body correlations in magic-angle twisted bilayer graphene},
\href{https://www.nature.com/articles/s41586-019-1422-x}{Nature, {\bf{572}}, 101 (2019)}.
\bibitem{origintwistedbilayer}
H. Po, L. Zou, A. Vishwanath, and T. Senthil, {\em Origin of Mott Insulating Behavior and Superconductivity in Twisted Bilayer Graphene}, \href{https://journals.aps.org/prx/abstract/10.1103/PhysRevX.8.031089}{Phys. Rev. X {\bf8}, 031089 (2018)}.

\bibitem{vanderwaals}
A. Geim and I. Grigorieva, {\em Van der Waals heterostructures}, \href{https://www.nature.com/articles/nature12385}{Nature {\bf 499}, 419 (2013)}.
\bibitem{twistedbilayermodel1}
R. Bistritzer and A.H. MacDonald, {\em Moir\'e bands in twisted double-layer graphene}, \href{https://www.pnas.org/content/108/30/12233}{PNAS {\bf108} (30) 12233-12237 (2011)}.
\bibitem{twistedbilayermodel2}
J. M. B. Lopes dos Santos, N. M. R. Peres, and A. H. Castro Neto, {\em Continuum model of the twisted graphene bilayer}, \href{https://journals.aps.org/prb/abstract/10.1103/PhysRevB.86.155449}{Phys. Rev. B {\bf 86}, 155449 (2012)}. 
\bibitem{twistedbilayermodel3}
J. M. B. Lopes dos Santos, N. M. R. Peres, and A. H. Castro Neto, {\em Graphene Bilayer with a Twist: Electronic Structure}, \href{https://journals.aps.org/prl/abstract/10.1103/PhysRevLett.99.256802}{Phys. Rev. Lett. {\bf 99}, 256802 (2007)}.
\bibitem{alexpaper}
G. Tarnopolsky, A.J. Kruchkov, and A. Vishwanath, {\em Origin of Magic Angles in Twisted Bilayer Graphene}, \href{https://journals.aps.org/prl/abstract/10.1103/PhysRevLett.122.106405}{Phys. Rev. Lett. {\bf122}, 106405. (2019)}
\bibitem{futwisted}
M. Koshino, N.F.Q. Yuan, T. Koretsune, M. Ochi, K. Kuroki, and L. Fu, {\em Maximally Localized Wannier Orbitals and the Extended Hubbard Model for Twisted Bilayer Graphene}, \href{https://journals.aps.org/prx/abstract/10.1103/PhysRevX.8.031087}{Phys. Rev. X {\bf8}, 031087 (2018)}.
\bibitem{originofhbngap1}
J.Jung, A.M. DaSilva, A.H. MacDonald and S. Adam, {\em Origin of band gaps in graphene on hexagonal boron nitride}, \href{https://www.nature.com/articles/ncomms7308}{Nature Communications, {\bf6}, 6308 (2015)}.
\bibitem{originofhbngap2}
P. San-Jose, A. Gutiérrez-Rubio, M. Sturla and F. Guinea, {\em Spontaneous strains and gap in graphene on boron nitride}, \href{https://journals.aps.org/prb/abstract/10.1103/PhysRevB.90.075428}{Phys. Rev. B {\bf 90}, 075428 (2014)}.
\bibitem{originofhbngap3}
F. Amet, J. R. Williams, K. Watanabe, T. Taniguchi, and D. Goldhaber-Gordon, {\em Insulating Behavior at the Neutrality Point in Single-Layer Graphene}, \href{https://journals.aps.org/prl/abstract/10.1103/PhysRevLett.110.216601}{Phys. Rev. Lett. {\bf 110}, 216601 (2013)}.
\bibitem{originofhbngap4}
B. Hunt, J. D. Sanchez-Yamagishi, A. F. Young,
M. Yankowitz, B. J. LeRoy, K. Watanabe, T. Taniguchi,
P. Moon, M. Koshino, P. Jarillo- Herrero, and R. C.
Ashoori, {\em Massive Dirac Fermions and Hofstadter Butterfly in a van der Waals Heterostructure}, \href{https://science.sciencemag.org/content/340/6139/1427}{Science {\bf 340}, 1427–1430 (2013)}.
\bibitem{Sharpe605}
A.L. Sharpe, E.J. Fox, A.W. Barnard, J. Finney, K. Watanabe,
T. Taniguchi, M.A. Kastner, and D. Goldhaber-Gordon,
{\em Emergent ferromagnetism near three-quarters filling in twisted
	bilayer graphene},
\href{https://science.sciencemag.org/content/365/6453/605}{Science, {\bf 365}, 605 (2019)}.
\bibitem{serlin2019intrinsic}
M. Serlin, C.L. Tschirhart, H. Polshyn, Y. Zhang, J. Zhu, K. Watanabe, T. Taniguchi,
L. Balents, and A.F. Young,
{\em Intrinsic quantized anomalous hall effect in a moir\'e
	heterostructure},
\href{https://arxiv.org/abs/1907.00261}{arXiv:1907.00261}.
\bibitem{chen2019evidence}
G. Chen, L. Jiang, S. Wu, B. Lyu, H. Li, B.L 
Chittari, K. Watanabe, T. Taniguchi, Z. Shi, J. Jung, et~al,
{\em Evidence of a gate-tunable mott insulator in a trilayer graphene
	moir{\'e} superlattice}, 
\href{https://www.nature.com/articles/s41567-018-0387-2}{ Nature Physics, {\bf 15}, 237 (2019)}.

\bibitem{Chen2019}
G. Chen, A.L. Sharpe, P. Gallagher, I.T. Rosen, E.J. Fox,
L. Jiang, B. Lyu, H. Li, K. Watanabe, T. Taniguchi, J. Jung, Z. Shi, D. Goldhaber-Gordon, Y. Zhang, and F. Wang,
{\em Signatures of tunable superconductivity in a trilayer graphene
	moir{\'e} superlattice.}
\href{https://www.nature.com/articles/s41586-019-1393-y}{Nature, {\bf 572} 215 (2019)}.
\bibitem{chen2019tunable}
G.Chen, A.L. Sharpe, E.J. Fox, Y. Zhang, S. Wang, L. Jiang,
B. Lyu, H. Li, K. Watanabe, T. Taniguchi, et~al,
{\em Tunable correlated chern insulator and ferromagnetism in trilayer
	graphene/boron nitride Moir\'e superlattice},
\href{https://arxiv.org/abs/1905.06535}{arXiv:1905.06535}.
\bibitem{repellin2019ferromagnetism}
C. Repellin, Z. Dong, Y.H. Zhang, and T. Senthil,
{\em Ferromagnetism in narrow bands of moir\'e superlattices},
\href{https://arxiv.org/abs/1907.11723}{arXiv:1907.11723}.
\bibitem{PhysRevLett.122.016401}
B.L. Chittari, G. Chen, Y. Zhang, F. Wang, and J. Jung,
{\em Gate-tunable topological flat bands in trilayer graphene
	boron-nitride moir\'e superlattices},
\href{https://journals.aps.org/prl/abstract/10.1103/PhysRevLett.122.016401}{Phys. Rev. Lett., {\bf 122}, 016401 (2019)}.

\bibitem{PhysRevB.99.075127}
Y.H. Zhang, D. Mao, Y. Cao, P. Jarillo-Herrero, and T. Senthil,
{\em Nearly flat chern bands in moir\'e superlattices},
\href{https://journals.aps.org/prb/abstract/10.1103/PhysRevB.99.075127}{ Phys. Rev. B, {\bf 99}, 075127 (2019)}.
\bibitem{senthilbridging}
Y.H. Zhang and T. Senthil, {\em Bridging Hubbard model physics and quantum Hall physics in trilayer graphene /h-BN moir\'e superlattice}, \href{https://journals.aps.org/prb/abstract/10.1103/PhysRevB.99.205150}{ Phys. Rev. B, {\bf 99}, 205150 (2019)}.
\bibitem{Constantinprb}
C. Schrade and L. Fu, {\em Spin-valley density wave in moiré materials}, \href{https://journals.aps.org/prb/abstract/10.1103/PhysRevB.100.035413}{Phys. Rev. B {\bf 100}, 035413 (2019).}
\bibitem{PhysRevB.82.035409}
F. Zhang, B. Sahu, H. Min, and A.H. MacDonald,
{\em Band structure of $abc$-stacked graphene trilayers},
\href{https://journals.aps.org/prb/abstract/10.1103/PhysRevB.82.035409}{Phys. Rev. B, {\bf82}, 035409 (2010)}.
\bibitem{2019arXiv190108209Z}
Y.H {Zhang}, D. {Mao}, and T.~{Senthil},
{\em Twisted Bilayer Graphene Aligned with Hexagonal Boron Nitride:
	Anomalous Hall Effect and a Lattice Model},
\href{https://journals.aps.org/prresearch/abstract/10.1103/PhysRevResearch.1.033126}{Phys. Rev. Research {\bf 1}, 033126 (2019)}.

\bibitem{zalatelanamolous}
N. Bultinck, S. Chatterjee, and M.P. Zaletel, {\em  Anomalous Hall ferromagnetism in twisted bilayer graphene}, \href{https://arxiv.org/abs/1901.08110}{arXiv:1901.08110}.
\bibitem{ladoprl}
T.M.R Wolf, J.L. Lado, G. Blatter and O. Zilberberg,{\em Electrically Tunable Flat Bands and Magnetism in Twisted Bilayer Graphene}, \href{https://journals.aps.org/prl/abstract/10.1103/PhysRevLett.123.096802}{Phys. Rev. Lett. {\bf 123}, 096802. (2019)}
\bibitem{supple}
See the supplementary materials for technical details of the single-particle Hamiltonians of TLG-hBN and TBG-hBN, the projected interaction, the Hartree-Fock corrections to the single-hole energy, comparing the bandwidth of the single-hole dispersion with the flat valence band dispersion and further numerical data about fractional Chern insulators.
\bibitem{lauchli2013hierarchy}
A.M. L{\"a}uchli, Z. Liu, E.J. Bergholtz, and R. Moessner.
{\em Hierarchy of fractional chern insulators and competing compressible
	states},
\href{https://journals.aps.org/prl/abstract/10.1103/PhysRevLett.111.126802}{ Phys. Rev. Lett. {\bf 111},126802 (2013)}.
\bibitem{Emilreview}
E.J. Bergholtz and Z. Liu,
{\em Topological Flat Band Models and Fractional Chern Insulators},
\href{http://www.worldscientific.com/doi/abs/10.1142/S021797921330017X}{Int. J. Mod. Phys. B {\bf 27}, 1330017 (2013)}.
\bibitem{fciprx}
N. Regnault and B. A. Bernevig, {\em Fractional Chern Insulator}, \href{https://doi.org/10.1103/PhysRevX.1.021014}{Phys. Rev. X {\bf 1}, 021014 (2011)}.
\bibitem{manybodysymmetry}
B. A. Bernevig and N. Regnault, {\em Emergent many-body translational symmetries of Abelian and non-Abelian fractionally filled topological insulators}, \href{https://doi.org/10.1103/PhysRevB.85.075128}{Phys. Rev. B {\bf 85}, 075128 (2012)}.
\bibitem{haldanestatistics}
Y.-L. Wu, N. Regnault, and B. A. Bernevig, {\em Haldane statistics for fractional Chern insulators with an arbitrary Chern number}, \href{https://doi.org/10.1103/PhysRevB.89.155113}{Phys. Rev. B {\bf 89}, 155113 (2014)}.
\bibitem{LiH}
H. Li and F. D. M. Haldane, {\em Entanglement Spectrum as a Generalization of Entanglement Entropy: Identification of Topological Order in Non-Abelian Fractional Quantum Hall Effect States}, \href{https://doi.org/10.1103/PhysRevLett.101.010504}{Phys. Rev. Lett. {\bf 101}, 010504 (2008)}.
\bibitem{PES} 
A. Sterdyniak, N. Regnault, and B. A. Bernevig, {\em Extracting Excitations from Model State Entanglement}, \href{https://doi.org/10.1103/PhysRevLett.106.100405}{Phys. Rev. Lett. {\bf 106}, 100405 (2011)}.
\bibitem{abinitiomoire}
J. Jung, A. Raoux, Z. Qiao, and A. H. MacDonald, {\em Ab initio theory of moire superlattice bands in layered two-dimensional materials}, \href{https://journals.aps.org/prb/abstract/10.1103/PhysRevB.89.205414}{Phys. Rev. B {\bf89}, 205414 (2014)}.
\end{thebibliography}
\end{document}